\documentclass[aps,prl,groupedaddress]{revtex4}

\usepackage{ulem}
\usepackage[usenames]{color}

\begin{document}


\title{Density functional scheme for calculating the pair density}


\author{Masahiko Higuchi}
\affiliation{Department of Physics, Faculty of Science, Shinshu University, 
Matsumoto 390-8621, Japan}

\author{Katsuhiko Higuchi}
\affiliation{Graduate School of Advanced Sciences of Matter, 
Hiroshima University, Higashi-Hiroshima 739-8527, Japan}


\date{\today}

\begin{abstract}
The density functional scheme for calculating the pair density 
is presented by means of the constrained-search 
technique. The resultant single-particle equation takes the form of the 
modified Hartree-Fock equation which contains the kinetic contribution of 
the exchange-correlation energy functional as the correlation potential. The 
practical form of the kinetic contribution is also proposed with the aid of 
the scaling relations of the kinetic energy functionals.
\end{abstract}


\maketitle

The pair density (PD) functional theory is expected to improve upon the 
conventional density functional theory\cite{1,2}, 
because the PD has a larger amount of information than the 
electron density\cite{3,4,5}.  
Recently the PD functional theory 
has been developed by several workers. 
Ziesche first proposed the PD functional theory using the natural 
spin geminals\cite{6,7}. Another scheme was offered by Gonis with taking 
the two-particle densities as a basic variable\cite{8,9}. Nagy generalized 
Gonis's theory to the PD functional theory 
which has the auxiliary equation of a two-particle problem\cite{10,11}.
Ayers presented the well-conceived approximate form of the kinetic energy 
functional within the PD functional theory\cite{12}. 

In this paper, we propose the other type of the PD functional 
theory. Our strategy is to reproduce the PD by using the single 
Slater determinant (SSD). 
This strategy has two kinds of merits. The first merit 
is that the reproduced PD is never unphysical. This is a strong 
merit because the $N$-representability of the PD has been an 
unresolved problem, as yet\cite{4,13,14,15,16,17,18,19}. The details 
will be discussed later. The second merit is the interesting form of the 
single-particle equation. As shown later, it is the modified Hartree-Fock 
equation which additionally contains the kinetic contribution of the 
exchange-correlation energy functional as the correlation term. This form is 
reasonably intelligible to us because correlation effects are explicitly 
incorporated into the single-particle equation as the correction to the 
Hartree-Fock equation. Hereafter, we shall describe our scheme and the above 
merits in detail.

Let us begin with the Hohenberg-Kohn theorems with taking the PD 
as a basic variable. The PD is defined by
\begin{eqnarray}
\label{eq1}
\gamma^{(2)}\!({\rm {\bf r{r}'}};{\rm {\bf r{r}'}}) 
=
\left\langle \Psi \right| \frac{1}{2}
\int\!\!\!\int {\hat {\psi }^{+}(x)\hat {\psi }^{+}(x')
\hat {\psi }(x')\hat {\psi }(x)} 
\rm{d}\eta \rm{d}{\eta }'\left| \Psi \right\rangle,
\end{eqnarray}
where $\hat {\psi }(x)$ and $\hat {\psi }^{+}(x)$ are 
field operators of electrons, $\Psi $ is the anitisymmetric wave function, 
and $x$ denotes the coordinates including 
the spatial coordinate $\rm {\bf r}$ and spin coordinate $\eta $. 
We consider a system, the Hamiltonian of which is given by
\begin{equation}
\label{eq2}
\hat {H}=\hat {T}+\hat {W}+\int {\hat {\rho }({\rm {\bf r}})v_{\rm{ext}} 
({\rm {\bf r}})\rm{d}{\rm {\bf r}}} ,
\end{equation}
where $\hat {T}$, $\hat {W}$ and $\hat {\rho }({\rm {\bf r}})$ are operators 
of the kinetic energy, electron-electron interaction and electron density, 
respectively, and $v_{\rm{ext}} ({\rm {\bf r}})$ stands for the external 
potential. The proof procedure is similar to the constrained-search 
theories\cite{20,21,22,23}. The universal functional 
$F[\gamma ^{(2)}]$ is defined by
\begin{eqnarray}
\label{eq3}
 F[\gamma ^{(2)}]
 &=&\mathop {\rm{Min}}\limits_{\Psi \to \gamma ^{(2)}} 
 \left\langle \Psi \right|\hat {T}+\hat {W}\left| \Psi \right\rangle 
 \nonumber \\ 
 &=&
 \left\langle {\Psi [\gamma ^{(2)}]} \right|\hat {T}+\hat {W}\left| {\Psi 
[\gamma ^{(2)}]} \right\rangle 
\end{eqnarray}
where $\Psi \to \gamma ^{(2)} $ 
denotes the searching over all anitisymmetric wave functions that yield a 
particular $\gamma ^{(2)} ({\rm {\bf r{r}'}};{\rm {\bf r{r}'}})$. In the 
second line, the minimizing wave function is expressed as $\Psi [\gamma 
^{(2)}]$. Here we suppose that the minimum exists in Eq. (\ref{eq3}).  
By virtue of the Rayleigh-Ritz principle, 
we get\cite{6,7}
\begin{equation}
\label{eq4}
\Psi [\gamma _{0}^{(2)} ]=\Psi _{0} \, ,
\end{equation}
where $\Psi _{0} $ 
is the ground-state wave function, and 
$\gamma _{0}^{(2)} ({\rm {\bf r{r}'}};{\rm {\bf r{r}'}})$ 
is the ground-state PD given by 
$\gamma _{0}^{(2)} ({\rm {\bf r{r}'}};{\rm {\bf r{r}'}})=
\left\langle {\Psi _{0} } \right|\hat {\gamma }^{(2)} 
({\rm {\bf r{r}'}};{\rm {\bf r{r}'}})\left| {\Psi _{0} } \right\rangle $. 


The variational principle with respect to $\gamma ^{(2)} ({\rm {\bf 
r{r}'}};{\rm {\bf r{r}'}})$ can be obtained by dividing the ordinary 
Rayleigh-Ritz principle into two steps: 
$E{ }_{0}=\mathop {\rm{Min}}\limits_{\gamma ^{(2)} } \left\{ {\mathop 
{\rm{Min}}\limits_{\Psi \to \gamma ^{(2)} } \left\langle \Psi \right|\hat 
{H}\left| \Psi \right\rangle } \right\}$.  
Using Eqs. (\ref{eq2}) and (\ref{eq3}), and defining 
the energy functional $E[\gamma ^{(2)} ]$ as 
\begin{equation}
\label{eq5}
E[\gamma ^{(2)} ]\equiv F[\gamma ^{(2)} ]+\frac{2}{N-1}\int\!\!\!\int 
{\gamma ^{(2)} ({\rm {\bf r{r}'}};{\rm {\bf r{r}'}})v_{\rm{ext}} ({\rm 
{\bf r}})\rm{d}{\rm {\bf r}}\rm{d}{\rm {\bf {r}'}}} ,
\end{equation}
the Rayleigh-Ritz principle is further rewritten as\cite{6,7}
\begin{equation}
\label{eq6}
\begin{array}{c}
 E{ }_{0}=\mathop {\rm{Min}}\limits_{\gamma ^{(2)} } E[\gamma ^{(2)} ] 
\\ 
 =E[\gamma _{0}^{(2)} ]. \\ 
 \end{array}
\end{equation}
Equations (\ref{eq4}) and (\ref{eq6}) are regarded as the extened 
Hohenberg-Kohn theorems. 

In order to perform the variational principle Eq. (\ref{eq6}), 
it is necessary to restrict 
the searching area within the set of $N$-representable PD's. 
Only the necessary 
condition for the $N$-representable PD, 
which is called Pauli condition, has 
been realized by Coleman\cite{13}. However, the necessary and sufficient 
condition for the $N$-representable PD is, as yet, the unresolved 
problem\cite{4,13,14,15,16,17,18,19}. 
Our strategy is to restrict the searching 
area within 
not the set of $N$-representable PD's 
but that of PD's which are given by SSD's. 
If the former and latter sets are respectively 
denoted as $C$ and ${C}'$, then $C \supseteq {C}'$ holds obviously. 
Therefore, this strategy prevents the minimizing PD from being unphysical 
one. There is a possibility that the ground-state PD 
may belong to $C \cap 
{\bar {C}}'$, where ${\bar {C}}'$ stands for the complementary set of ${C}'$. 
However, our strategy is to search the best solution 
within ${C}'$. Best solution means the most appropriate reproduction of the 
PD with respect to searching the minimum of Eq. (\ref{eq6}) within ${C}'$. 
This spirit is analogous to the usual Hartree-Fock approximation because 
the best antisymmetric wave function is searched within the set 
of SSD's\cite{24}. Along this spirit, we shall give the practical and 
useful scheme for calculating the PD. 

In order to perform the variational principle practically, we shall 
introduce the kinetic energy functional $T_{s} [\gamma ^{(2)}]$ which is 
defined by
\begin{equation}
\label{eq7}
T_{s} [\gamma ^{(2)}]\equiv \mathop {\rm{Min}}\limits_{\Phi \to \gamma 
^{(2)}} \left\langle \Phi \right|\hat {T}\left| \Phi \right\rangle \,,
\end{equation}
where $\Phi $ is the SSD 
which is constructed 
from $N$ orthonormal spin orbitals $\psi _{\mu } (x)$, and 
$\Phi \to \gamma ^{(2)}$ indicates that the search is constrained among all 
SSD's which yield the prescribed $\gamma ^{(2)}({\rm {\bf r{r}'}};{\rm 
{\bf r{r}'}})$. We again suppose that the minimum exists in 
Eq. (\ref{eq7}) similarly to Eq. (\ref{eq3}). In Eq. (\ref{eq7}), 
$\gamma ^{(2)}({\rm {\bf r{r}'}};{\rm 
{\bf r{r}'}})$ is expressed as the expectation value with respect to the 
SSD. Using Eq. (\ref{eq1}), we have
\begin{eqnarray}
\label{eq8}
\!\!\!\!\!
\gamma ^{(2)}({\rm {\bf r{r}'}};{\rm {\bf r{r}'}})
 =\frac{1}{2} \sum\limits_{\mu ,\,\nu =1}^N 
{\int\!\!\!\int {\left\{ 
{\psi _{\mu }^{\ast } (x)\psi _{\nu }^{\ast } (x')
\psi _{\mu } (x)\psi _{\nu } (x')} \right.} } -\left. 
{\psi _{\mu }^{\ast } (x)\psi _{\nu }^{\ast } (x')
\psi _{\nu } (x)\psi _{\mu } (x')} \right\} \rm{d}\eta \rm{d}{\eta }'. 
\end{eqnarray}
Taking the minimization of Eq. (\ref{eq7}) by the use of 
the Lagrange method of undetermined multipliers, the equation for 
minimizing spin orbitals can be obtained as follows:
\begin{eqnarray}
\label{eq9}
-\frac{\hbar ^{2}\nabla ^{2}}{2m}\psi _{\xi } (x) + 
\int {\rm{d}} x'\frac{\mu ({\rm {\bf r}},{\rm {\bf {r}'}}) +
\mu ({\rm {\bf {r}'}},{\rm {\bf r}})}{2} 
\sum \limits_{{\xi }'=1}^N \{ \psi_{{\xi }'}^{\ast} (x') 
\psi_{{\xi}'} (x') 
\psi_{\xi} (x) \!-\!
\psi_{{\xi }'}^{\ast} (x')\psi_{\xi} (x')\psi_{{\xi}'} (x) \} \nonumber \\
=\sum \limits_{\nu =1}^N {\varepsilon} _{{\xi} {\nu} }  {\psi}_{\nu} (x),
\end{eqnarray}
where $\mu ({\rm {\bf r}},\,{\rm {\bf {r}'}})$ and $\varepsilon _{\xi \nu } 
$ are the Lagrange multipliers which respectively 
correspond to the restriction 
of Eq. (\ref{eq8}) and orthonormality of spin orbitals. 
The Lagrange multiplier 
function $\mu ({\rm {\bf r}},\,{\rm {\bf {r}'}})$ should be determined by 
requiring spin orbitals to yield a given $\gamma ^{(2)} ({\rm {\bf 
r{r}'}};{\rm {\bf r{r}'}})$. That is, $\mu ({\rm {\bf r}},\,{\rm {\bf 
{r}'}})$ can be written as $\mu ({\rm {\bf r}},\,{\rm {\bf {r}'}})=\mu 
[\gamma ^{(2)}]({\rm {\bf r}},\,{\rm {\bf {r}'}})$. 

Using $F[\gamma ^{(2)}]$ and $T_{s} [\gamma ^{(2)} ]$, the 
exchange-correlation energy functional 
$E_{xc} [\gamma ^{(2)}]$ is defined as
\begin{equation}
\label{eq10}
F[\gamma ^{(2)}]=T_{s} [\gamma ^{(2)}]+U[\gamma ^{(2)}]+
E_{xc} [\gamma ^{(2)}],
\end{equation}
where $U[\gamma ^{(2)}]$ is the usual Hartree term. Here we need the note on 
Eq. (\ref{eq10}). The domain of $F[\gamma ^{(2)}]$ is $N$-representable PD's 
, that is the set $C$.
On the other hand, 
$T_{s} [\gamma ^{(2)} ]$ is defined with reference to PD's 
which belong to the set $C'$. 

Substituting Eq. (\ref{eq10}) into Eq. (\ref{eq5}) and 
taking the variation with respect to the PD, 
we can get 
$\mu [\tilde{\gamma} _{0}^{(2)} ]
({\rm {\bf r}},\,{\rm {\bf {r}'}})$ 
of Eq. (\ref{eq9}), which reproduces 
$\tilde{\gamma} _{0}^{(2)} ({\rm {\bf r{r}'}};{\rm {\bf r{r}'}})$:
\begin{eqnarray}
\label{eq11}
\mu [\tilde{\gamma} _{0}^{(2)} ]
({\rm {\bf r}},\,{\rm {\bf {r}'}}) 
=\left. \frac{\delta U[\gamma ^{(2)}]}{\delta \gamma ^{(2)}
({\rm {\bf r{r}'}};{\rm {\bf r{r}'}})} 
\right |_{\gamma^{(2)}=\tilde{\gamma} _{0}^{(2)} }  
+\left.
\frac{\delta E_{xc} [\gamma ^{(2)}]}{\delta \gamma ^{(2)}
({\rm {\bf r{r}'}};{\rm {\bf r{r}'}})} 
\right|_{\gamma^{(2)}=\tilde{\gamma} _{0}^{(2)}}  
+\frac{2}{N-1}v_{ext} ({\rm {\bf r}}),
\end{eqnarray}
where 
$\tilde{\gamma} _{0}^{(2)} ({\rm {\bf r{r}'}};{\rm {\bf r{r}'}})$ is
the best solution within $C'$.

Next we consider the exchange-correlation energy functional 
$E_{xc} [\gamma ^{(2)}]$. Using Eq. (\ref{eq1}), Eq. (\ref{eq3}) 
is rewritten as
\begin{equation}
\label{eq12}
F[\gamma ^{(2)}]=T[\gamma ^{(2)}]+e^{2}\int\!\!\!\int {\rm{d}{\rm {\bf 
r}}\rm{d}{\rm {\bf {r}'}}} \frac{\gamma ^{(2)}({\rm {\bf r{r}'}};{\rm {\bf 
r{r}'}})}{\left| {{\rm {\bf r}}-{\rm {\bf {r}'}}} \right|}\quad ,
\end{equation}
where $T[\gamma ^{(2)}]$ is defined as the expectation value of $\hat {T}$ 
with respect to the wave function $\Psi [\gamma ^{(2)}]$. Note that 
$T[\gamma ^{(2)}]$ coincides with the ground-state value of the real kinetic 
energy when $\gamma ^{(2)}=\gamma _{0}^{(2)} $. Comparing Eq. (\ref{eq12}) 
to Eq. (\ref{eq10}), we get
\begin{equation}
\label{eq13}
E_{xc} [\gamma ^{(2)}]=\Delta T_{xc}
 [\gamma ^{(2)}]+{E}'_{xc} [\gamma ^{(2)}],
\end{equation}
where the kinetic contribution of the exchange-correlation energy functional 
$\Delta T_{xc} [\gamma ^{(2)}]$ and the authentic 
exchange-correlation energy 
${E}'_{xc} [\gamma ^{(2)}]$ are respectively given by
\begin{equation}
\label{eq14}
\Delta T_{xc} [\gamma ^{(2)}]\equiv 
T[\gamma ^{(2)}]-T_{s} [\gamma ^{(2)}],
\end{equation}
and
\begin{equation}
\label{eq15}
{E}'_{xc} [\gamma ^{(2)}]=\frac{e^{2}}{2}\int\!\!\!\int {\rm{d}{\rm {\bf 
r}}\rm{d}{\rm {\bf {r}'}}} \frac{2\gamma ^{(2)}({\rm {\bf r{r}'}};{\rm {\bf 
r{r}'}})}{\left| {{\rm {\bf r}}-{\rm {\bf {r}'}}} \right|}-
U[\gamma ^{(2)}].
\end{equation}
Equation (\ref{eq13}) is similar to that of the conventional DFT.  
Here note that devising the approximate form of 
${E}'_{xc} [\gamma ^{(2)}]$ is not 
needed in the present scheme since ${E}'_{xc} [\gamma ^{(2)}]$ is explicitly 
expressed as a functional of the PD. 

Substitution of Eqs. (\ref{eq11}) and (\ref{eq13}) into Eq. (\ref{eq9}), 
followed by the unitary 
transformation of $\varepsilon _{\mu \nu } $ and constant terms which are 
brought by the substitution, leads to the single-particle equation of the 
canonical form;
\begin{widetext}
\begin{eqnarray}
\label{eq16}
& &\left\{ {-\frac{\hbar ^{2}\nabla ^{2}}{2m}+v_{ext} ({\rm {\bf r}})} 
\right\}\chi _{\xi } (x) 
\nonumber \\ 
& & +
\int  {{\rm{d}} x'  \left\{  {\frac{e^{2}}{\left| 
{{\rm {\bf r}} - {\rm {\bf {r}'}}} \right|}
+ \frac{1}{2}\frac{\delta \Delta T_{xc} [\gamma ^{(2)}]}
{\delta \gamma ^{(2)}({\rm {\bf r{r}'}};{\rm {\bf r{r}'}})}
\Bigg|_{\gamma^{(2)}=\tilde{\gamma}_{0} ^{(2)}}
+ \frac{1}{2}\frac{\delta \Delta T_{xc} [\gamma ^{(2)}]}
{\delta \gamma ^{(2)}({\rm {\bf {r}'r}};{\rm {\bf {r}'r}})}}
\Bigg|_{\gamma^{(2)}=\tilde{\gamma}_{0} ^{(2)}}
 \right\}} 
\nonumber \\ 
& & \,\,\,\,\,\,\,\,\,\,\,\,\,\,\,\,\,\,\,\,\,\,\,\,\,\,\,\,\,\,\,\,\,\,\,
\times
\sum\limits_{\nu =1}^N  {\left\{ 
 \chi _{\nu }^{\ast } (x')
 \chi _{\nu } (x')
 \chi _{\xi } (x)
\!-\!
 \chi _{\nu }^{\ast } (x')
 \chi _{\xi } (x')
 \chi _{\nu } (x) 
\right\}} 
=
\varepsilon _{\xi } \chi _{\xi } (x ),
\end{eqnarray}
\end{widetext}
where $\chi _{\xi } (x)$ is the spin orbital which is 
converted from $\psi _{\xi } (x)$ via the unitary 
transformation. Since the SSD is generally kept invariant under the unitary 
transformation, the reproduced form of the PD 
of Eq. (\ref{eq8}) 
is rewritten as 
\begin{eqnarray}
\label{eq17}
\!\!\!\!\!\!\!\!
\tilde{\gamma}_{0} ^{(2)} ({\rm {\bf r{r}'}};{\rm {\bf r{r}'}})
=\frac{1}{2}
\sum\limits_{\mu ,\,\nu =1}^N 
{\int\!\!\!\int {\left\{ 
{\chi _{\mu }^{\ast } (x)
\chi _{\nu }^{\ast } (x')
\chi _{\mu } (x )
\chi _{\nu } (x')} 
\right.} } 
\!-\! \left. 
{\chi _{\mu }^{\ast } (x)
\chi _{\nu }^{\ast } (x')
\chi _{\nu } (x )
\chi _{\mu } (x')} 
\right\} \rm{d}\eta \rm{d}{\eta }'.
\end{eqnarray}
Equations (\ref{eq16}) and (\ref{eq17}) are final expressions for 
the single-particle 
equations which should be solved self-consistently. It is free from the 
self-interaction like the Hartree-Fock equation.

The single-particle equation (\ref{eq16}) is intelligible to us 
because it can be 
regarded as the modified Hartree-Fock equation to which kinetic 
contributions are added. If $\Delta T_{xc} [\gamma ^{(2)} ]$ terms are 
neglected, Eq. (\ref{eq16}) exactly coincides with the Hartree-Fock 
equation. Due to 
$\Delta T_{xc} [\gamma ^{(2)} ]$ terms, spin orbitals deviate from 
solutions of the Hartree-Fock equation.  Not only the additional terms 
concerning $\Delta 
T_{xc} [\gamma ^{(2)} ]$, but also values of the Hartree and Fock 
potentials are different from the Hartree-Fock equation. All of these 
differences come from $\Delta T_{xc} [\gamma ^{(2)} ]$ terms, and 
should be recognized as correlation effects. 

From another point of view, let 
us show again that $\Delta T_{xc} [\gamma ^{(2)} ]$ terms are the cause 
of correlation effects in the present scheme. Along the argument in 
Ref. \cite{22}, the following functional is introduced here: 
\begin{eqnarray}
\label{eq18}
{e}'[\Phi ]=\left\langle \Phi \right|\hat {T}\left| \Phi \right\rangle 
+e^{2}\int\!\!\!\int {\frac{\gamma ^{(2)}
({\rm {\bf r{r}'}};{\rm {\bf r{r}'}})}
{\left| {{\rm {\bf r}}-{\rm {\bf {r}'}}} \right|}\rm{d}{\rm {\bf r}}
\rm{d}{\rm {\bf {r}'}}} 
+\frac{2}{N-1}\int {v_{ext} ({\rm {\bf r}})
\gamma ^{(2)}({\rm {\bf r{r}'}};{\rm {\bf r{r}'}})
\rm{d}{\rm {\bf r}}\rm{d}{\rm {\bf {r}'}}} ,
\end{eqnarray}
where $\gamma ^{(2)}({\rm {\bf r{r}'}};{\rm {\bf r{r}'}})$ is expressed as 
the expectation value with respect to the SSD $\Phi $. 
${e}'\,[\Phi ]$ can be 
rewritten as the expectation value of $\hat {H}$ with respect to $\Phi $. 
Therefore, the minimum value of ${e}'\,[\Phi ]$ is obtained at $\Phi =\Phi 
_{HF} $, where $\Phi _{HF} $ is the ground-state wave function of the 
Hartree-Fock approximation. On the other hand, we can also show that 
$\mathop {\rm{Min}}\limits_\Phi {e}'\,[\Phi ]=\mathop 
{\rm{Min}}\limits_{\gamma ^{(2)}\in {C}'} \left\{ {E\,[\gamma ^{(2)} 
]-\Delta T_{xc} \,[\gamma ^{(2)} ]} \right\}$ similarly to the proof of 
Eq. (4-12) of Ref. \cite{22}. Therefore, we get
\begin{equation}
\label{eq19}
\left\langle {\Phi _{HF} } \right|\hat {H}\left| {\Phi _{HF} } \right\rangle 
=\mathop {\rm{Min}}\limits_{\gamma ^{(2)}\in {C}'} \left\{ {E\,[\gamma 
^{(2)}]-\Delta T_{xc} \,[\gamma ^{(2)}]} \right\}.
\end{equation}
This relation indicates that the ground-state energy which is expressed as 
$\mathop {\rm{Min}}\limits_{\gamma ^{(2)}\in {C}'} E[\gamma ^{(2)}]$ in our 
scheme is different from that calculated in the Hartree-Fock approximation. 
The difference originates from $\Delta T_{xc} [\gamma ^{(2)}]$. Since 
$\Delta T_{xc} [\gamma ^{(2)}]$ is rewritten as 
$\mathop {\rm{Min}}\limits_{\Psi \to \gamma^{(2)}} \left\langle 
\Psi \right|\hat {T}\left| \Psi \right\rangle -
\mathop {\rm{Min}}\limits_{\Phi \to \gamma^{(2)}} \left\langle 
\Phi \right|\hat {T}\left| \Phi \right\rangle $
\cite{26}, it holds the inequality 
$\Delta T_{xc} [\gamma ^{(2)}] \le 0$. 
Therefore, 
it is expected that the ground-state energy becomes lower than $\left\langle 
{\Phi _{HF} } \right|\hat {H}\left| {\Phi _{HF} } \right\rangle $. That is 
to say, correlation effects are incorporated into the present scheme if 
$\Delta T_{xc} [\gamma ^{(2)}]$ is taken into account. 

The negative value of 
$\Delta T_{xc} [\gamma^{(2)} ]$ 
also leads to the 
inequality 
$T_{s} [\gamma^{(2)} ] \ge T_{HF} $\cite{27}, 
where $T_{HF} $ is the kinetic energy within the Hartree-Fock approximation. 
This means again 
that correlation effects are incorporated into the single-particle equation 
through $\Delta T_{xc} [\gamma ^{(2)}]$ because correlation effects have a 
tendency to raise the kinetic energy. In addition, we shall give a 
comment on the possibility of $\Delta T_{xc} [\gamma ^{(2)}] = 0$. 
If the wave functions which yield the PD of $C'$ were SSD's alone, 
$\Delta T_{xc} [\gamma ^{(2)}]$ would be always zero. 
However, that is not the case. 
Indeed, there exists the set of wave functions which give 
the PD of $C'$ and are not expressed by a SSD\cite{19}. 
Therefore, it is not the case 
that $\Delta T_{xc} [\gamma ^{(2)}]$ is always zero. 

In order to make it more feasible to perform the actual calculation of the 
present scheme, we propose the approximate form of $\Delta T_{xc} [\gamma 
^{(2)}]$ by using the scaling relation of $\Delta T_{xc} [\gamma ^{(2)}]$ as 
a sum rule. The scaling relation of $\Delta T_{xc} [\gamma ^{(2)}]$ can be 
obtained from those of $T[\gamma ^{(2)}]$ and $T_{s} [\gamma ^{(2)}]$. As 
for the scaling relation of $T[\gamma ^{(2)}]$, 
Levy and Ziesche\cite{28} has derived the relation 
$T[\gamma _{\zeta }^{(2)} ]=\zeta ^{2}T[\gamma ^{(2)}]$. Here 
$\gamma _{\zeta }^{(2)} ({\rm {\bf r{r}'}};{\rm {\bf r{r}'}})$ denotes the 
expectation value of $\hat {\gamma }^{(2)}({\rm {\bf r{r}'}};{\rm {\bf 
r{r}'}})$ with respect to the scaled wave function, i.e., 
$\gamma _{\zeta }^{(2)} ({\rm {\bf r{r}'}};{\rm {\bf r{r}'}})=
\zeta ^{6}\gamma ^{(2)}(\zeta {\rm {\bf r}}\,\zeta {\rm {\bf {r}'}};\zeta 
{\rm {\bf r}}\,\zeta {\rm {\bf {r}'}})$.
In the similar way to deriving the above relation, the scaling relation of 
$T_{s} [\gamma ^{(2)} ]$ 
is also given by 
$T_{s} [\gamma _{\zeta }^{(2)} ]=\zeta ^{2}T_{s} [\gamma ^{(2)} ]$\cite{29}. 
Thus, we get 
$\Delta T_{xc} [\gamma _{\zeta }^{(2)} ]=\zeta ^{2}\Delta T_{xc} 
[\gamma ^{(2)} ].$
Along the technique of Ref. \cite{30}, let 
$\mathop {\lim }\limits_{\zeta \to 1} \frac{\partial }{\partial \zeta }$ 
act on both sides of this relation. Utilizing 
the integration by parts, we obtain\cite{31}
\begin{eqnarray}
\label{eq20}
2\Delta T_{xc} [\gamma ^{(2)}]\!=\!
&-&
\int\!\!\!\int
{\gamma ^{(2)}({\rm {\bf r{r}'}};{\rm {\bf r{r}'}})} {\rm {\bf r}}\!
\cdot \!\! \nabla \left\{ {\frac{\delta 
\Delta T_{xc} [\gamma ^{(2)}]}{\delta \gamma ^{(2)}({\rm {\bf r{r}'}};{\rm 
{\bf r{r}'}})}} \right\}\rm{d}{\rm {\bf r}}\rm{d}{\rm {\bf {r}'}} \nonumber \\
&-&
\int\!\!\!\int {\gamma ^{(2)}({\rm {\bf r{r}'}};{\rm {\bf r{r}'}})} 
{\rm {\bf r}}\cdot \nabla \left\{ {\frac{\delta \Delta T_{xc} [\gamma 
^{(2)}]}{\delta \gamma ^{(2)}({\rm {\bf {r}'r}};{\rm {\bf {r}'r}})}} 
\right\}\rm{d}{\rm {\bf r}}\rm{d}{\rm {\bf {r}'}}.
\end{eqnarray}
We shall use this equation as a sum rule for 
$\Delta T_{xc} [\gamma ^{(2)}]$. 

With reference to the local density approximation of the 
conventional DFT\cite{2}, the following form 
is assumed for $\Delta T_{xc} [\gamma ^{(2)}]$: 
\begin{equation}
\label{eq21}
\Delta T_{xc} [\gamma ^{(2)} ]=\int\!\!\!\int {\left. {\Delta t_{xc} 
(\gamma ^{(2)} )} \right|_{\gamma ^{(2)} =\gamma ^{(2)} ({\rm {\bf 
r{r}'}};{\rm {\bf r{r}'}})} } \rm{d}{\rm {\bf r}}\rm{d}{\rm {\bf {r}'}},
\end{equation}
where $\Delta t_{xc} (\gamma ^{(2)} )$ is an ordinary function of 
$\gamma ^{(2)} $. 
Substituting Eq. (\ref{eq21}) into Eq. (\ref{eq20}), and neglecting 
the surface integral at infinity, 
we get the differential equation with respect to 
$\Delta t_{xc} (\gamma ^{(2)} )$ as the necessary condition:
\begin{equation}
\label{eq22}
4\Delta t_{xc} (\gamma ^{(2)} )-3\gamma ^{(2)} \frac{\partial \Delta 
t_{xc} (\gamma ^{(2)} )}{\partial \gamma ^{(2)} }=0.
\end{equation}
A similar relation for $\Delta t_{xc} (\rho )$ has been derived by Levy and 
Perdew in the conventional DFT\cite{32}. Solving the differential equation 
(\ref{eq22}), and substituting the solution into Eq. (\ref{eq21}), 
we finally get the 
approximate form of $\Delta T_{xc} [\gamma ^{(2)} ]$;
\begin{equation}
\label{eq23}
\Delta T_{xc} [\gamma ^{(2)} ]=K\int\!\!\!\int {\gamma ^{(2)} ({\rm 
{\bf r{r}'}};{\rm {\bf r{r}'}})^{\frac{4}{3}}\rm{d}{\rm {\bf r}}\rm{d}{\rm 
{\bf {r}'}}} ,
\end{equation}
where $K$ is the arbitrary constant. The constant $K$ may be available as an 
adjustable parameter in our 
scheme. Namely, $K$ may be determined 
corresponding to the magnitude of correlation effects of the system. For 
example, in case of atomic structures, we may fix the value by fitting 
total energies to those of more accurate calculations\cite{33}. 

In conclusion, this paper is aimed at searching the best solution of the 
PD within the set of $C'$. 
The biggest merit of our theory is that 
the solution is physically reasonable. Namely, our solution is necessarily 
$N$-representable. The solution even in the set $C'$ 
can contain the correlation 
effects through $\Delta T_{xc}$ term more or less but definitely. This point 
is also important. It seems worthwhile to perform the actual calculation so 
as to check how much the present scheme can reflect correlation effects 
beyond the Hartree-Fock. This is the next step to be done\cite{33}. 


\begin{thebibliography}{00}
\bibitem{27} 
The correlation effects likely raise the kinetic energy, i.e., 
$T[\gamma^{(2)} ] \ge T_{HF} $. 
Since $\Delta T_{xc} [\gamma^{(2)}] \le 0$, 
we obtain the inequality 
$T_{s} [\gamma^{(2)} ] \ge T_{HF} $.
\bibitem{31} 
It is easily confirmed that the same relation as Eq. (\ref{eq20}) can be 
obtained for $T[\gamma ^{(2)} ]$ by employing the virial theorem. As for 
the virial theorem, see, Ref. \cite{29}.
\end{thebibliography}
\end{document}